\documentclass{elsarticle}
\usepackage{graphics}
\usepackage{amssymb}
\usepackage{amsmath}
\usepackage{wasysym}
\usepackage{latexsym}
\usepackage{graphicx}
\usepackage{epsfig}
\usepackage{subfigure}
\usepackage{float}
\usepackage{color}
\usepackage{lineno}
\usepackage{verbatim}
\usepackage{enumitem}
\usepackage[T1]{fontenc}

\journal{Physica A: Statistical Mechanics and its Applications}
\begin{document}

\begin{frontmatter}

\title{When Less is More: Evolutionary Dynamics of Deception in a Sender-Receiver Game}

\author[mymainaddress]{Eduardo V.  M. Vieira}
\ead{eduardov@usp.br}

\author[mymainaddress]{Jos\'e F.  Fontanari\corref{mycorrespondingauthor}}
\cortext[mycorrespondingauthor]{Corresponding author}
\ead{fontanari@ifsc.usp.br}

\address[mymainaddress]{Instituto de F\'{\i}sica de S\~ao Carlos, Universidade de S\~ao Paulo, 13560-970 S\~ao Carlos, S\~ao Paulo, Brazil}

\begin{abstract}
The spread of disinformation poses a significant threat to societal well-being.  We analyze this phenomenon using an evolutionary game theory model of the sender-receiver game, where senders aim to mislead receivers and receivers aim to discern the truth.   Using a combination of replicator equations, finite-size scaling analysis, and extensive Monte Carlo simulations, we investigate the long-term evolutionary dynamics of this game.  Our central finding is a counterintuitive threshold phenomenon: the role (sender or receiver) with the larger difference in payoffs between successful and unsuccessful interactions is surprisingly more likely to lose in the long run.  We show that this effect is robust across different parameter values and arises from the interplay between the relative speeds of evolution of the two roles and the ability of the slower evolving role to exploit the fixed strategy of the faster evolving role. Moreover, for finite populations we find that the initially less frequent strategy of the slower role is more likely to fixate in the population. The initially rarer strategy in the less-rewarded role is, paradoxically, more likely to prevail.
\end{abstract}


\end{frontmatter}

\section{Introduction}\label{sec1}

The struggle between truth and falsehood is a defining characteristic of the human condition, a tension explored by philosophers for millennia.  Plato's Ring of Gyges \cite{Plato_2007} and Kant's analysis of promising \cite{Kant_2012} (see also \cite{Sober_1994,Fontanari_2023}) highlight the enduring nature of this conflict.  Yet, while these philosophical inquiries delve into the moral dimensions of lying, the phenomenon itself extends beyond human ethics, appearing in the animal kingdom in the evolution of signaling and mimicry \cite{Maynard_2003,Zahavi_1975}.  Today, this ancient struggle takes on a new urgency with the breakdown of epistemic security, fueled by the rapid spread of disinformation (i.e., misinformation with the explicit intent to mislead \cite{Fallis_2015})  through social media and deepfake technology.  This erosion of trust undermines our ability to address pressing societal issues, from climate change to public health \cite{Seger_2020}.  Here we explore the conflict between lying and telling the truth using evolutionary game theory \cite{Maynard_1982} to  determine which role (sender or receiver) is more likely to win  a sender-receiver game \cite{Kreps_1994} where the sender is rewarded for misleading the receiver and the receiver is rewarded for getting the correct information.

While behavioral economics experiments offer valuable insights into honesty and deception \cite{Erat_2012}, they typically involve transparent payoff structures where all participants are aware of the consequences of both truthful and deceptive actions. This full information setting, while useful for studying individual behavior, may not fully capture the dynamics of disinformation, where the intent to mislead is often obscured.  To better model this aspect, we consider a simplified sender-receiver game where the sender's goal is to manipulate the receiver's belief, and the receiver aims to infer the true state of the world.  Specifically, the sender, instead of rolling a die as in the original  behavioral economics experiments \cite{Erat_2012}, flips a coin and reports the outcome to the receiver, who must then decide whether to believe the report. This simplification, by removing the element of chance, allows us to focus on the strategic interaction between sender and receiver and analyze the evolutionary pressures favoring truth or deception. Senders can either report the true coin flip or lie, while receivers can choose to believe or disbelieve.  Importantly, in our evolutionary framework, players only have access to the payoffs of others playing the same role (sender or receiver), mimicking the information asymmetry often present in real-world disinformation scenarios.

The simplified setup of the sender-receiver game, in which the original random dice roll is replaced by a coin toss, captures the core antagonistic interaction between senders and receivers and focuses on the fundamental evolutionary dynamics of deception.  This inherent antagonism is somewhat obscured by the significant random component introduced by guessing the outcome of a dice roll. In particular, a disbeliever may not receive the true outcome even when the sender is lying (the probability of guessing correctly is 1/5), while a believer always receives the correct outcome when the sender is truthful. The different types of lies (e.g., `black' or `white') considered in the  behavioral economics experiments \cite{Erat_2012} are effectively determined by the payoff matrix, and can therefore  be studied within the simplified coin toss scenario. However, in the context of deception and misinformation, `black' lies -- those that harm a believing receiver -- are of primary concern.

Evolutionary game theory is particularly well-suited for this study because it allows us to model the interaction between senders and receivers as a dynamic, adaptive process. It accounts for the fact that senders and receivers continuously adjust their strategies in response to each other, based on their relative success. This dynamic interplay is crucial for understanding the spread and persistence of misinformation and deception. Specifically, it enables us to examine the long-term consequences of different strategies and identify stable or unstable outcomes, providing insights into the persistence or collapse of deceptive behaviors. Moreover, evolutionary game theory has a long-standing history in the study of the evolution of deception. As noted by Sober, this framework provides a powerful tool for analyzing the adaptive nature of lying \cite{Sober_1994}. Furthermore, the original behavioral experiments on the evolution of lying  \cite{Erat_2012} were designed to facilitate imitation between subjects, a scenario subsequently analyzed using an evolutionary game theory formulation \cite{Capraro_2019,Capraro_2020}. These studies highlight the relevance and applicability of evolutionary game theory to understanding the dynamics of deception.

The infinite-population limit of this asymmetric sender-receiver game, where sender and receiver interests are directly opposed, has been extensively studied using the replicator equation \cite{Gaunersdorfer_1991,Hofbauer_1998}.  This approach predicts persistent oscillations in strategy frequencies, reflecting the absence of stable equilibria, as any strategy can be exploited by a counter-strategy in the other role. However, these oscillations are not absorbing states under stochastic imitation dynamics, which models finite populations and converges to the replicator equation in the infinite-population limit \cite{Traulsen_2005,Ohtsuki_2006,Fontanari_2024b}. In this imitation dynamics, players adopt strategies from those with higher payoffs, with the probability of imitation increasing linearly with the payoff difference.  Crucially, the dynamics converge to one of four absorbing states, corresponding to the fixation of a single strategy  within each role. 
One example of an absorbing state is the fixation of liars in the sender population and believers in the believer population, resulting in consistent wins for the senders.  We focus on the probability of fixation on each absorbing state and the mean fixation time. This focus on absorbing states offers a distinct perspective, revealing dynamics drastically different from the oscillatory behavior predicted by the replicator equation, even when considering the infinite-population limit of the stochastic model.

To investigate the stochastic dynamics of finite populations, we employ extensive Monte Carlo simulations \cite{Newman_1999,Landau_2000} combined with finite-size scaling analysis \cite{Privman_1990}. Use of Monte Carlo and finite-size scaling in conjunction is a common practice in statistical physics, especially when the interest is to infer the behavior of infinitely large systems by simulating  small systems \cite{Privman_1990}. For smaller populations, these simulations are validated by numerical solutions of the corresponding birth-death process \cite{Karlin_1975}.  Our key finding is that the mean fixation time scales linearly with population size. This implies that the stochastic dynamics quickly transitions away from the oscillatory regime predicted by the replicator equation and rapidly converges to one of the absorbing states. This contrasts sharply with the typical exponential scaling of the mean time to escape coexistence equilibria \cite{Antal_2006,Fontanari_2024a}.  Furthermore, we identify a threshold in the payoff parameter space that separates regions where senders or receivers are more likely to achieve fixation.  We find that the role with less at stake - or, equivalently, the role that evolves more slowly - is more likely to win the game.   While the initially rarer strategy within the slower evolving role has a higher probability of fixation, this effect diminishes as population size increases. Surprisingly, imitation dynamics favor the fixation of the strategy with lower growth rate and initial frequency, contradicting the typical outcome of replicator dynamics for biological macromolecules \cite{Eigen_1971,Mariano_2024}.

The remainder of this paper is structured as follows. Section \ref{sec:srg} defines the payoff matrices for the sender-receiver game, specifying the sender's intent to mislead. Section \ref{sec:id} describes the stochastic imitation dynamics, where players adopt strategies from higher-payoff individuals within their role, with imitation probability increasing linearly with payoff difference.  For completeness, Section \ref{sec:re}  reviews the replicator equation and its oscillatory solutions in the infinite-population limit. Section \ref{sec:mc} presents Monte Carlo simulations of the stochastic dynamics, focusing on fixation probabilities and mean fixation times, and employing finite-size scaling to extrapolate to the infinite-population limit.  Section \ref{sec:bd} uses a birth-death process formulation to  derive the equations for the fixation probabilities and mean  fixation times,  and validate the Monte Carlo results for small populations.  Finally, Section \ref{sec:disc} offers concluding remarks.

\section{The sender-receiver game}\label{sec:srg}

The sender-receiver game \cite{Kreps_1994} involves two roles, each with at least two strategies.  We consider a simplified version where $N$ players are permanently assigned as senders and $N$ as receivers \cite{Gaunersdorfer_1991}, unlike versions where roles are randomly assigned each round \cite{Silva_2009,Vieira_2024}.  Our fixed-role game is defined as follows.

The sender privately observes the result of a coin toss and then chooses whether to report the truth or lie.  Both options are costless, and the sender's payoff depends on whether the receiver believes the report, according to the sender's payoff matrix
\begin{equation}\label{sender}
\begin{array}{lcc}
& \mbox{Believe} & \mbox{Disbelieve} \\
 \mbox{Lie} & b_s & -c_s \\
\mbox{Tell the truth}& -c_s & b_s
\end{array} 
\end{equation}
where  $b_s$ and $c_s$ are non-negative parameters. This payoff matrix shows that the sender aims to deceive the receiver, gaining a reward only if the receiver incorrectly guesses the coin toss outcome. Conversely, the receiver is rewarded for correctly guessing the outcome and penalized for guessing incorrectly,  according to the receiver's payoff matrix
\begin{equation}\label{receiver}
\begin{array}{lcc}
 & \mbox{True} & \mbox{False} \\
\mbox{Believe} & b_r & -c_r \\
\mbox{Disbelieve}& -c_r & b_r 
\end{array} 
\end{equation}
where, as before,  $b_r$ and $c_r$ are non-negative parameters.  Evolving Batesian mimicry, like the viceroy butterfly mimicking the poisonous monarch, can be modeled by a variant this game  \cite{Sober_1994}.  The viceroy butterflies are the senders, their blue jays predators are the receivers, and their interests are opposed, as shown in the payoff matrices.

 \section{Imitation dynamics}\label{sec:id}

Consider a well-mixed population of $X$ liars and $Y$ believers at time $t$.  The numbers of truth tellers and disbelievers are $N-X$ and $N-Y$, respectively. At each time step $\delta t$, a focal sender $i_s$ and a focal receiver $i_r$ are randomly chosen and play a round of the sender-receiver game, receiving payoffs $f_{i_s}$ and $f_{i_r}$ according to payoff matrices (\ref{sender}) and (\ref{receiver}).  A model sender $j_s \neq i_s$ and a model receiver $j_r \neq i_r$ are then randomly chosen, and they also play a round of the game, resulting in payoffs $f_{j_s}$ and $f_{j_r}$.  Focal individuals only update their strategies by imitating more successful peers; thus, $i_s$ and $i_r$ do not change strategies if $f_{j_s} \leq f_{i_s}$ and $f_{j_r} \leq f_{i_r}$. However, when  $f_{j_s} > f_{i_s}$,  the  probability that the focal sender $i_s$  switches to the strategy of the model sender $j_s$  is
\begin{equation}\label{probs}
 \frac{f_{j_s} - f_{i_s}}{\Delta f_{max}}.
\end{equation}
Similarly, when  $f_{j_r} > f_{i_r}$,  the  probability that the focal receiver $i_r$  switches to the strategy of the model receiver $j_r$  is
\begin{equation}\label{probr}
 \frac{f_{j_r} - f_{i_r}}{\Delta f_{max}} .
\end{equation}
Here  
\begin{equation}\label{Dmax}
\Delta f_{max} = \max(b_s+c_s,b_r+c_r)
\end{equation}
guarantees that the probabilities (\ref{probs}) and (\ref{probr})  are not greater than $1$.   Although senders  $i_s$ and  $j_s$ might use the same strategy, their payoffs can vary (e.g.,  $f_{j_s} > f_{i_s}$) because they interact with different receivers. If $i_s$ were to imitate $j_s$ in this scenario, it would not alter the population composition.  Crucially, given that the numerators of (\ref{probs}) and (\ref{probr}) are  $b_s+c_s$ and $b_r+c_r$, respectively, the imitation process depends only on the ratio $(b_r+c_r)/(b_s+c_s)$.

After the attempted strategy update, the time step $\delta t$ ends, and the time variable $t$ is updated to $t + \delta t$. The simulation continues until the stochastic dynamics converge to an absorbing state where both sender and receiver populations are homogeneous.
 For fixed $t$, the switching probabilities  (\ref{probs}) and (\ref{probr}), which permit copying only from more successful individuals, yield the replicator equation in the large-population limit $N \to \infty$  when the  time step is  $\delta t = 1/N$ \cite{Fontanari_2024b}. 
 For switching probabilities that are linear with respect to payoff differences and recover the replicator equation in the large-population limit, see  \cite{Traulsen_2005,Ohtsuki_2006}.
  
In the following analysis, we examine the imitation dynamics of the sender-receiver game within both the deterministic regime ($N \to \infty$) and finite population settings, employing Monte Carlo simulations and an analytical formulation based on the birth-death process.
 
\section{Replicator equation approach}\label{sec:re}

The replicator equation provides a deterministic description of the competition between strategies in the sender-receiver game, specifically in the limit of infinitely many senders and receivers ($N \to \infty$).  We define $x$ as the proportion of liars and $1-x$ as the proportion of truth tellers among senders. Similarly, $y$ represents the proportion of believers, with $1-y$ representing the proportion of incredulous receivers.

We derive the replicator equations for the coupled dynamics of $x$ and $y$ by calculating the mean payoff of each strategy.  Among senders, the mean payoff for liars is $ \pi_l = b_s y - c_s (1-y)$, while the mean payoff for truth tellers is $\pi_t = b_s (1-y) - c_s y$. The average payoff for senders is  $\bar{\pi}_s = x \pi_l + (1-x) \pi_t$. 
The replicator equation describes a competitive process where strategies proliferate if their mean payoff is greater than the population's mean payoff \cite{Hofbauer_1998}, i.e.
\begin{eqnarray}\label{x_rep}
\Delta f_{max} \frac{dx}{dt} &  = &  x (\pi_l - \bar{\pi}_s )  \nonumber \\
&  = &  x (1-x) (\pi_l - \pi_t)  \nonumber \\
&  = &  (b_s+c_s) x (1-x) (2y-1) .
\end{eqnarray}
A similar analysis for the receiver population leads to
\begin{equation}\label{y_rep}
\Delta f_{max} \frac{dy}{dt} = - (b_r+c_r) y (1-y) (2x-1) .
\end{equation}
 While the factor  $\Delta f_{max} $ in these equations is crucial for linking them to the deterministic limit of the stochastic imitation process  \cite{Traulsen_2005,Ohtsuki_2006,Fontanari_2024b},  it only influences the timescale. Equation (\ref{x_rep}) indicates that the liar population grows when believers constitute more than half the population. Conversely, eq. (\ref{y_rep}) shows that the believer population shrinks when liars make up more than half the population.

The equilibrium solutions  of the replicator equations  (\ref{x_rep}) and    (\ref{y_rep}), denoted by  $x^*$ and $y^*$, are obtained by setting $dx/dt=dy/dt=0$. Their  local stability is determined  by linearizing these equations at the equilibrium solutions, resulting in the linear system 
\begin{equation}\label{linear}
\begin{pmatrix} 
\frac{du}{dt} \\ 
 \frac{dv}{dt}  
\end{pmatrix} = \mathbf{A}
\begin{pmatrix} 
u \\ 
 v   
 \end{pmatrix} ,
\end{equation}
where $u=x-x^*$, $v=y-y^*$,  and 
\begin{equation}
 \mathbf{A} = \frac{1}{\Delta f_{max} }
 \begin{pmatrix} 
(b_s+c_s) (1-2x^*)(2y^*-1) & 2(b_s+c_s)x^*(1-x^*)\\ 
-2(b_r+c_r)y^*(1-y^*) & -(b_r+c_r)(1-2y^*)(2x^*-1)
\end{pmatrix}
\end{equation}
is the Jacobian or community matrix  evaluated at the equilibrium solutions.  
Local stability of the equilibrium solutions is determined by the signs of the real parts of the eigenvalues \cite{Britton_2003,Murray_2007}.  We briefly describe these equilibria below:
 \begin{enumerate}[label=(\alph*)]

\item The equilibrium $x^* =y^*=0$ corresponds to a sender population consisting only of truth tellers and a receiver population consisting only of disbelivers.  The eigenvalues are $\lambda_s = -(b_s + c_s)/\Delta f_{max}$  and  $\lambda_r = (b_r + c_r)/\Delta f_{max}$ which indicate that this  equilibrium is a saddle point.  This result reflects the fact that such population cannot be invaded by liars but can be invaded by believers. 

\item  The equilibrium $x^* = 0$ and $y^*=1$ corresponds to a sender population consisting only of truth tellers and a receiver population consisting only of believers. The eigenvalues are $\lambda_s = (b_s + c_s)/\Delta f_{max}$  and  $\lambda_r = -(b_r + c_r)/\Delta f_{max}$ so this  equilibrium is  also a saddle point: the population can be invaded by liars but cannot be invaded by disbelivers. 

\item  The equilibrium $x^* = 1$ and $y^*=0$ corresponds to a sender population consisting only of liars and a receiver population consisting only of disbelivers. The eigenvalues are $\lambda_s = (b_s + c_s)/\Delta f_{max}$  and  $\lambda_r = -(b_r + c_r)/\Delta f_{max}$ so this  equilibrium is  also a saddle point: the population can be invaded by  truth tellers  but cannot be invaded by believers. 

\item  The equilibrium $x^* = 1$ and $y^*=1$ corresponds to a sender population consisting only of liars and a receiver population consisting only of believers. The eigenvalues are $\lambda_s = -(b_s + c_s)/\Delta f_{max} $  and  $\lambda_r = (b_r + c_r)/\Delta f_{max}$ so this  equilibrium is  also a saddle point: the population cannot be invaded by  truth tellers  but can be invaded by disbelivers. 

\item  The equilibrium $x^* = y^* = 1/2$ corresponds to the coexistence of all strategies. The eigenvalues are the complex conjugates
\begin{eqnarray}\label{autoi}
\lambda_s & = &  \frac{\mathbf{i}}{2 \Delta f_{max}}  \sqrt{( b_s + c_s)( b_r + c_r)} \nonumber \\
\lambda_r & = &  - \frac{\mathbf{i}}{2 \Delta f_{max}}  \sqrt{( b_s + c_s)( b_r + c_r)} ,
 \end{eqnarray}
 where $\mathbf{i}$ is the imaginary unit. Since their real part is zero, this equilibrium is neutral.  

\end{enumerate}
 Since eqs. (\ref{x_rep}) and (\ref{y_rep}) have no stable equilibria, the solutions oscillate around the neutral fixed point  $x^* = y^* = 1/2$ \cite{Britton_2003,Murray_2007}.

The Hamiltonian that characterizes the game dynamics can be written as \cite{Hofbauer_1998}
\begin{equation}
 H(x,y)=  y \left(1-y\right)x^{\gamma}\left(1-x\right)^{\gamma} ,
\end{equation}
where
\begin{equation}\label{gamma}
\gamma = \frac{b_r+c_r}{b_s+c_s} . 
\end{equation}
It can be shown that $dH/dt = 0$, so $H=C$ is a constant of motion. In fact, 
the closed trajectories in the  phase plane are the solutions of  the equation
\begin{equation}\label{orb0}
\frac{dy}{dx} = - \gamma \frac{y (1-y) (2x-1)}{ x (1-x) (2y-1)},
\end{equation}
which can be obtained by direct integration,
\begin{equation}\label{orb1}
y(1-y) [x(1-x)]^{\gamma } =  C, 
\end{equation}
where 
\begin{equation}\label{C}
C=y_0(1-y_0)[x_0 (1-x_0)]^\gamma \leq 1/4^{1+\gamma},
\end{equation}
 with $x_0 = x(0)$ and $y_0 = y(0)$.  The system is conservative and the orbit is determined by the initial conditions $x_0$ and $y_0$. 

\begin{figure}[th] 
\centering
 \includegraphics[width=1\columnwidth]{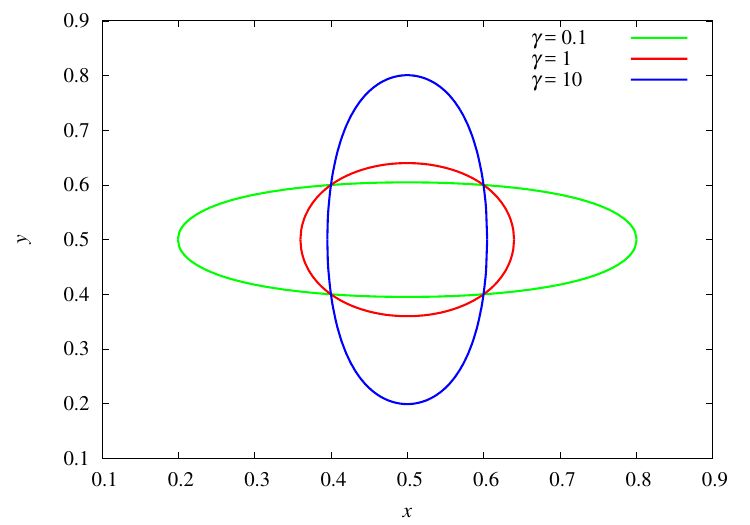}  
\caption{Phase plane trajectories showing liar frequency ($x$) and believer frequency ($y$) for  $\gamma = 0.1$, $1$, and $10$. Initial conditions: $x_0=y_0 = 0.6$. Trajectories are clockwise and centered at the neutral fixed point ($x^*=y^* = 1/2$).
 }  
\label{fig:1}  
\end{figure}

The orbit equation (\ref{orb1}) allows to determine the amplitude of the oscillations. For example, the condition $dy/dx=0$ together with the orbit equation gives the minimum $y_{min}$ and the maximum $y_{max}$ values that $y$ can take. We find
\begin{eqnarray}
y_{min} & =  & \frac{1}{2} - \frac{1}{2} \sqrt{1- 4^{1+ \gamma} C} \label{ymin} \\
y_{max} & =  & \frac{1}{2} + \frac{1}{2} \sqrt{1- 4^{1+ \gamma} C}  \label{ymax} .
\end{eqnarray}
A similar analysis leads to  the minimum $x_{min}$ and the maximum $x_{max}$ values that $x$ can take, viz.
\begin{eqnarray}
x_{min} & =  & \frac{1}{2} - \frac{1}{2} \sqrt{1- 4 (4C)^{1/\gamma}} \label{xmin}\\  
x_{max} & =  & \frac{1}{2} + \frac{1}{2} \sqrt{1- 4 (4C)^{1/\gamma}} \label{xmax} .
\end{eqnarray}
The phase plane trajectories exhibit symmetry with respect to the transformations $x\leftrightarrow y$ and  $\gamma \leftrightarrow 1/\gamma$.  Figure \ref{fig:1} illustrates three such trajectories, all starting from the same initial condition ($x_0=y_0=0.6$) but with varying values of $\gamma$. As predicted by eq.  (\ref{C}), these trajectories pass through the points  $(x_0,y_0)$, $(1-x_0,y_0)$, $(x_0,1-y_0)$, and $(1-x_0,1-y_0)$.

\begin{figure}[th] 
\centering
 \includegraphics[width=1\columnwidth]{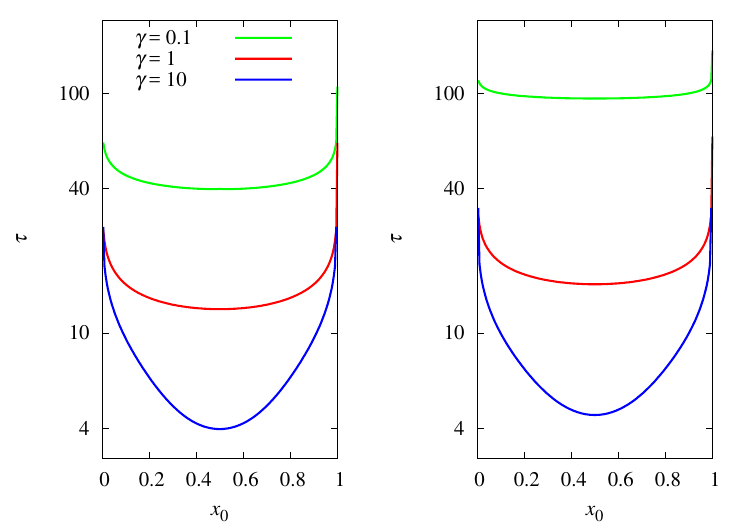}  
\caption{ Scaled period ($\tau$)  as a function of the initial liar frequency ($x_0$) for $\gamma = 0.1$, $1$, and $10$. The initial believer frequency is $y_0=0.5$ (left panel) and $y_0=0.1$ (right panel). Logarithmic scale on the y-axis for ease of comparison.
 }  
\label{fig:2}  
\end{figure}

Equations (\ref{ymin})-(\ref{xmax})  show that  oscillation amplitudes decrease as  $C \to 1/4^{1+\gamma}$, or equivalently,  as
$x_0 \to 1/2$ and $y_0 \to 1/2$, indicating that the orbits shrink towards the neutral fixed point $x^*=y^*=1/2$. These orbits of vanishingly small amplitudes  are governed by eq. (\ref{linear}),
\begin{eqnarray}
\Delta f_{max} \frac{du}{dt} & =  & \frac{1}{2} (b_s+c_s) v \\
\Delta f_{max} \frac{dv}{dt} & =  & - \frac{1}{2} (b_r+c_r) u,
\end{eqnarray}
which reduces to the equation of motion for the classical harmonic oscillator,
\begin{equation}
\frac{d^2u}{dt^2} + \frac{1}{4 [\Delta f_{max}]^2} (b_s+c_s)(b_r+c_r) u = 0.
\end{equation}
Thus the period of the small oscillations is 
\begin{equation}\label{Ts1}
T =  \frac{4\pi \Delta f_{max} }{\sqrt{ (b_s+c_s)(b_r+c_r)}} , 
\end{equation}
which could be obtained directly by  using  $T = 2\pi/\mbox{Im}(\lambda_s)$  where $\lambda_s$ is an eigenvalue of the community matrix for the neutral fixed point given in eq. (\ref{autoi}).
Introducing the scaled period
\begin{equation}\label{tau}
\tau = \frac{b_s+c_s}{\Delta f_{max}}  T, 
\end{equation}
 we rewrite eq. (\ref{Ts1}) as
\begin{equation}\label{Ts2}
\tau =  4\pi \gamma^{-1/2}, 
\end{equation}
which is  more convenient for comparison with the periods of finite amplitude oscillations.  

 For general initial conditions (i.e.,  $C$ not close to  its extreme value),  the period $T$ is calculated by using  eq. (\ref{orb1}) to eliminate $y$ from eq.  (\ref{x_rep}), resulting in
\begin{equation}
\frac{\Delta f_{max}}{b_s+c_s} \frac{dx}{dt} =  [x(1-x)]^{1-\gamma/2} \sqrt{ [x(1-x)]^\gamma -4 C}
\end{equation}
for $t \in [0,T/2]$, subject to the boundary conditions $x(0) = x_{min}$ and $x(T/2) = x_{max}$.
Integrating this equation yields
\begin{equation}\label{Tg1}
\tau = 2 \int_{x_{min}}^{x_{max}} \frac{dx}{[x(1-x)]^{1-\gamma/2} \sqrt{ [x(1-x)]^\gamma -4 C}}. 
\end{equation}
The integrand diverges  at the limits of integration because
$x_{min}(1-x_{min}) = x_{max}(1-x_{max}) =  (4C)^{1/\gamma}$.
\ref{ref:A} demonstrates the derivation of eq.  (\ref{Ts2}) from eq. (\ref{Tg1}) in the limit of small oscillations ($C \to  1/4^{1+\gamma}$).   \ref{ref:B} details the transformation of this improper integral into a proper integral suitable for numerical computation.

Figure \ref{fig:2} displays the scaled period as a function of the initial liar frequency for two distinct initial believer frequencies. Although our analytical expressions involve both $\gamma$ and $C$, the latter is a function of all three independent parameters ($x_0$, $y_0$, and $\gamma$) as  shown in eq. (\ref{C}), making $\gamma$ and $x_0$ the more informative variables.  The left panel ($y_0=0.5$) 
 illustrates the small oscillation limit ($x_0 \to 1/2$), where periods are given by eq. (\ref{Ts2}).  While the orbits for  $\gamma =0.1$ and $\gamma=10$ are symmetric (Fig. \ref{fig:1}), their corresponding periods differ considerably. This difference arises from the dependence of the rates of change of $x$ and $y$ on $\gamma$ given by eqs. (\ref{x_rep}) and (\ref{y_rep}).  
 When $ b_s + c_s < b_r + c_r$, the rate of change of $y$ is directly proportional to  $\gamma$, while the rate of change of $x$ depends on  $\gamma$ indirectly, via its influence on $y$. Conversely, when $ b_s + c_s > b_r + c_r$, the rate of change of $x$ is directly proportional to $\gamma$, while the rate of change of $y$ depends on $\gamma$ indirectly, via its influence on $x$. The period remains invariant under the transformations $x_0 \to 1-x_0$ and  $y_0 \to 1-y_0$  because $C$ is unchanged by these transformations.

Although the replicator equations predict intuitive oscillatory behavior, driven by the interplay of dominant and counter-strategies (e.g., lying and disbelieving), demographic noise destabilizes these solutions.  The noise drives the system towards absorbing states, which correspond to the saddle points discussed in our analysis of the equilibrium solutions.  

\begin{figure}[t] 
\centering
 \includegraphics[width=1\columnwidth]{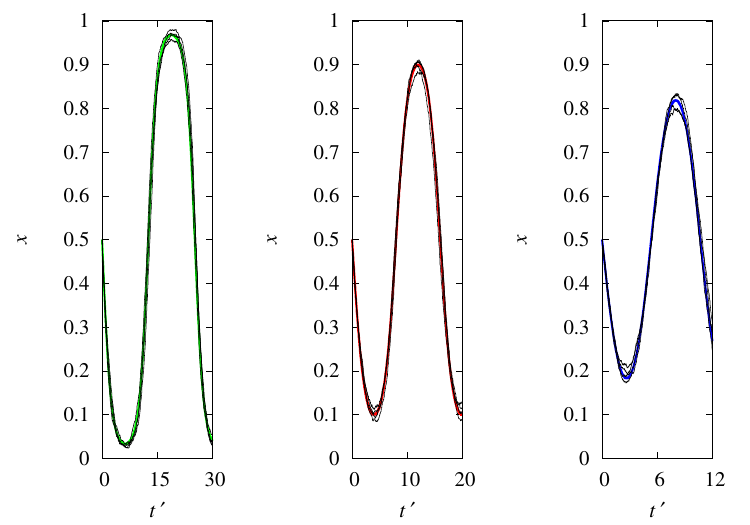}  
\caption{ Frequency  of liars as a function of the scaled time $t'= (b_s+c_s)t/\Delta f_{max}$ for   $\gamma = 0.5$ (left panel), $\gamma =1$ (middle panel) , and $\gamma =2$  (right panel). The initial frequencies of liars and believers are $x_0=0.5$ and $y_0=0.1$.   The thick colored curves are the numerical solutions of the replicator equations, and the thin black  curves are runs of the stochastic imitation dynamics with $N=1000$.
 }  
\label{fig:3}  
\end{figure}

\section{Monte Carlo simulations}\label{sec:mc}

\begin{figure}[t] 
\centering
 \includegraphics[width=1\columnwidth]{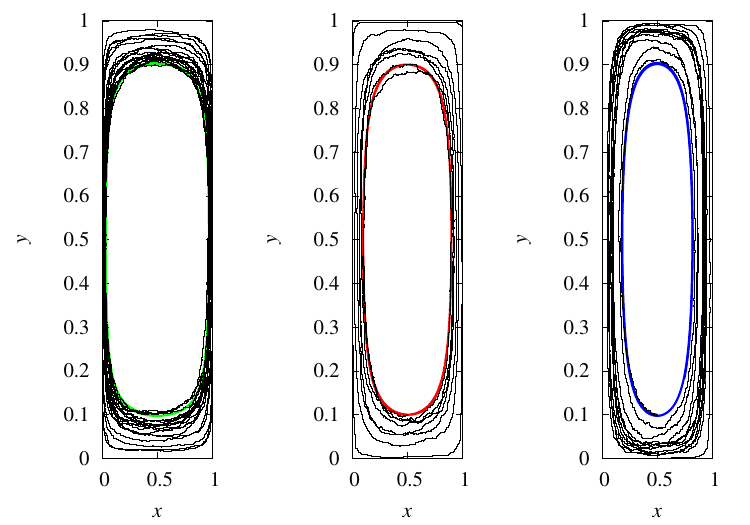}  
\caption{ Trajectories in the phase plane showing the frequency of liars $x$ and the frequency of believers $y$ for $\gamma = 0.5$ (left panel), $\gamma =1$ (middle panel) , and $\gamma =2$  (right panel).   The initial frequencies of liars and believers are  $x_0=0.5$ and  $y_0=0.1$.  The thick colored curves are the closed orbits given by eq. (\ref{orb1}), and the thin black  curves are runs of the stochastic imitation dynamics with $N=1000$. All orbits are in a clockwise direction. 
 }  
\label{fig:4}  
\end{figure}

Monte Carlo simulations provide a straightforward implementation of the imitation dynamics described in Section \ref{sec:id}. Figure \ref{fig:3} shows the close agreement between the deterministic predictions of the replicator equations and four  stochastic simulations  with $N=1000$ over short timescales (approximately one oscillation period). However,  over longer times,  the stochastic trajectories deviate from the deterministic orbits and converge to  absorbing states, as illustrated  in  Fig. \ref{fig:4}. For the particular  run shown in the panel with $\gamma = 0.5$, first the liars die out and then the believers take over the receiver population, leading to the absorbing state $x=0$ and $y=1$.  For the panel with  $\gamma=1$, the liars fixate first, followed by the fixation of the disbelievers, freezing the dynamics in the absorbing state $x=1$ and $y=0$. For the panel with $\gamma = 2$, the believers die out first, followed by the liars, leading to the absorbing state $x=0$ and $y=0$. Of course, other runs may lead to different absorbing states and our goal is to determine the probability of the dynamics reaching each of the four possible  absorbing states.

Let $p_{ld}$, $p_{lb}$, $p_{tb}$, and $p_{td}$ represent the probabilities of fixation for the four possible absorbing states: liars and disbelievers, liars and believers, truth tellers and believers, and truth tellers and disbelievers, respectively, such  that $p_{ld} + p_{lb} + p_{tb} + p_{td} = 1$.  These probabilities are estimated empirically from $10^5$ independent stochastic simulations for each parameter configuration. Because the imitation dynamics depend solely on the ratio  $\gamma$ given in eq. (\ref{gamma}), as discussed in Section \ref{sec:id}, the specific values of the parameters in the payoff matrices (\ref{sender}) and (\ref{receiver}) are not required.

\begin{figure}[t] 
\centering
 \includegraphics[width=1\columnwidth]{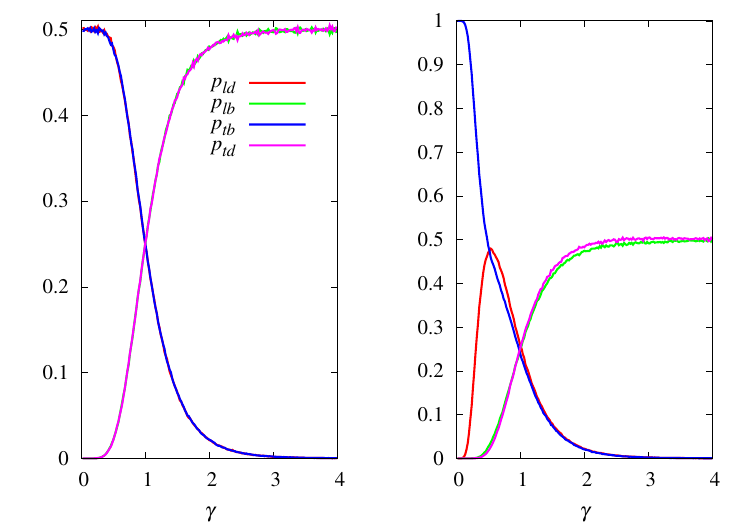}  
\caption{ Probability of fixation of the sender and receiver  strategies as a function of $\gamma$ for  $x_0=0.5$, $y_0=0.5$ (left panel), and $y_0=0.1$ (right panel).  The number of players in each role is   $N=100$.  In the symmetric scenario of the left panel we  have $p_{ld}=p_{tb}$  and $p_{lb}=p_{td}$.
 }  
\label{fig:5}  
\end{figure}

Figure \ref{fig:5} shows the fixation probabilities of the sender and receiver strategies for the initial frequency of liars $x_0=0.5$ and two values of the initial frequency of believers, $y_0=0.5$ and $y_0=0.1$. The former setting corresponds to a symmetric scenario where there is no difference between the absorbing state composed of liars and disbelivers and that  composed of truth tellers and believers, so that $p_{ld}=p_{tb}$.  In fact, both absorbing states correspond to the scenario where the receiver wins the sender-receiver game by correctly guessing the outcome of the coin toss.  Absorption into these states is more likely for $\gamma < 1$ (i.e., $\Delta f_{max} = b_s+c_s$), because a successful sender is copied with probability one, while a successful receiver is copied with probability $\gamma < 1$. This means that successful senders spread faster than successful receivers, so fixation is likely to occur first in the sender population. Then one of the receiver strategies will explore the sender's fixed strategy and eventually fixate.  This is exactly what happens in Fig. \ref{fig:4} for $\gamma=0.5$: first the truth tellers fixate, and then the believers quickly take over the receiver population, winning the game for the receivers. 
 Similar remarks apply to the equivalence between the 
 absorbing state consisting of liars and believers and the absorbing state consisting of truth tellers and disbelivers, so that $p_{lb}=p_{td}$. In this case, the sender wins the sender-receiver game by getting the receiver to incorrectly guess the outcome of the coin toss. As shown in Fig. \ref{fig:4} for $\gamma=2$, the receiver's defeat was sealed when the disbelivers fixated first.
As expected, $ p_{ld}=p_{tb}=p_{lb}=p_{td} = 0.25$ for $\gamma=1$.

Interestingly, in the case where the initial conditions favor the disbelivers (e.g., $y_0=0.1$) shown in the right panel of Fig. \ref{fig:5}, we find that believers are much more likely to fixate than disbelivers for small $\gamma$. In fact, in this regime, the initial condition strongly breaks the symmetry between the two  absorbing states  associated to the receiver win, giving $p_{tb} \gg p_{ld}$. This can be understood by noting that for small $\gamma$ the sender population evolves much faster than the receiver population, and the initial high frequency of disbelivers drives the rapid growth and fixation of the truth tellers, paving the way for  the fixation of the believers.  Except in the small $\gamma$ region, varying the initial frequency of believers $y_0$ has little  effect on the fixation probability of the different absorbing states. Figure \ref{fig:4} shows that this is to be expected, since if fixation of one of the strategies is not very fast, the stochastic trajectory will pass close to the antipode  believer frequency $1-y_0$, making the choice of a small initial $y_0$ irrelevant for the long-term outcome.  The results shown in both panels of Fig. \ref{fig:5} are explained by the surprising  result that the winner of the sender-receiver game is the role that evolves more slowly. In fact, in the sender-receiver game, the slowest role and the rarest strategy in each role are advantageous, which is the opposite of the outcome of the competition among molecular replicators \cite{Eigen_1971,Mariano_2024}.  

\begin{figure}[th] 
\centering
 \includegraphics[width=1\columnwidth]{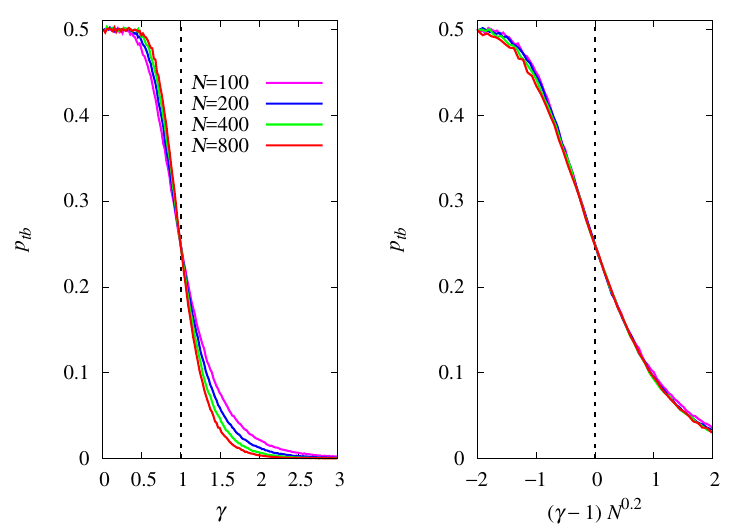}  
\caption{Probability of fixation of truth tellers and believers $p_{tb}$ as function of $\gamma$  (left panel) and as a function of the scaled variable $(\gamma-\gamma_c) N^{\alpha}$ with $\gamma_c=1$ and $\alpha = 0.2$ for  $N=100$, $200$, $400$ and $800$. The initial conditions are $x_0=y_0= 0.5$.
 }  
\label{fig:6}  
\end{figure}

For symmetric initial conditions $x_0=y_0=0.5$, Fig. \ref{fig:6} shows that increasing the population size $N$ leads to a sharp transition between the receiver-winner and the sender-winner regimes  at the threshold  $\gamma_c=1$.  More precisely, in the limit $N \to \infty$ we have $p_{tb}=p_{ld}=0.5$ for $\gamma < \gamma_c$ (receiver wins), $p_{tb}=p_{ld}=0$ for $\gamma > \gamma_c$ (sender wins), and $p_{tb}=p_{ld}=0.25$ for $\gamma = \gamma_c$.  Complementary results hold for the probability of fixation in the other  absorbing states.  The effect of  finite population (or demographic noise)  is  relevant only in the region close to the  threshold. In this region the scaling assumption  $p_{tb} =f [ (\gamma -\gamma_c) N^{\alpha}]$  perfectly describes the dependence on $\gamma$ and $N$,  as shown by the collapse of the curves for different population sizes  for $\gamma_c=1$ and $\alpha=0.2$. Here, $f(x)$ is a scaling function such that $f(x) \to 0$ when $x \to \infty$, $f(x) \to 0.5$ when $x \to -\infty$ and $f(0)=0.25$ (see \cite{Privman_1990,Kirkpatrick_1994,Campos_1999} for details and applications of the finite-size scaling analysis). The steepness of the threshold transition, given by the derivative of $p_{tb}$ with respect to $\gamma$ calculated at the threshold $\gamma_c$, increases with $N^{\alpha}$.  The small value of the exponent $\alpha$  is an indication of the weak effect of demographic noise on the fixation probabilities.

\begin{figure}[th] 
\centering
 \includegraphics[width=1\columnwidth]{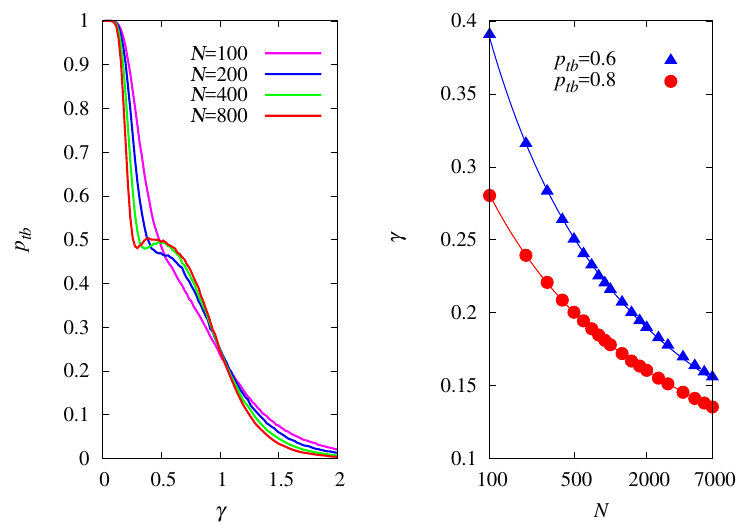}  
\caption{Probability of fixation of truth tellers and believers $p_{tb}$ as function of $\gamma$ for  $N=100$, $200$, $400$ and $800$ (left panel).  The right panel shows the values of $\gamma$ for  which  $p_{tb} =0.6$  and $p_{tb} =0.8$ as a function of $N$ on a logarithmic scale.  The curves are the fit function $\gamma = a/\ln N + b/\ln^2 N$, where $a$ and $b$ are fit parameters. The initial conditions are $x_0=0.5$ and $y_0= 0.1$. 
 }  
\label{fig:7}  
\end{figure}

However, the demographic noise has a much stronger effect in the case of non-symmetric initial conditions $x_0=0.5$ and $y_0=0.1$, shown in Fig. \ref{fig:7}. It turns out that the dominance of the rarest receiver strategy observed in Fig. \ref{fig:5} is a finite population phenomenon:  the boundary of  the region where $p_{tb} \approx 1$ approaches $\gamma=0$ very slowly as $N$ increases, disappearing completely in the limit $N \to \infty$.   To quantify this observation,   for fixed $N$ we  use a bisection method to determine the value of  $\gamma$ for which $p_{tb}=0.6$ and $0.8$, and  then we plot these values against $N$, as shown in the right panel of  Fig. \ref{fig:7}. Numerical fitting of the data shows that $\gamma \to 0$ as a power series of $1/\ln N$ for large $N$, with the term independent of $N$ set to zero.  For $N>100$ the first two non-zero terms of the series already fit the data very well as shown in the figure. This means that the effect of the initial conditions disappears for  large $N$: there will be no noticeable differences between the left panels of Figs. \ref{fig:6} and \ref{fig:7}, except in a region very close to $\gamma=0$, which shrinks as $1/\ln N$. In particular,  for large $N$ the threshold transition at $\gamma_c =1$ is identical to that for the symmetric initial conditions.  In fact, the diminishing influence of the initial conditions $x_0$ and $y_0$ with increasing $N$ is expected, since, as mentioned before, the larger $N$ the greater the chances that the stochastic dynamics reaches the antipode state $1-x_0$ and $1-y_0$ after closely following the deterministic orbits for about one period. This is the reason why in Fig. \ref{fig:3} we used the large population size $N=1000$ to test the agreement between the stochastic dynamics and the predictions of the replicator equations even for short times.
 
Besides the fixation probability, another important quantity to characterize the stochastic dynamics is the mean  fixation time $t_f$, i.e., the mean time for the dynamics to  freeze in an absorbing  state. In Section \ref{sec:bd} we will consider the conditional mean  fixation times, i.e., the mean time for the dynamics to freeze in a particular absorbing state.  Note that $t_f$ is dimensionless and is measured in Monte Carlo steps, where a Monte Carlo step corresponds to $N$ attempts to update randomly selected sender and receiver players, as described in Section \ref{sec:id}.   

\begin{figure}[th] 
\centering
 \includegraphics[width=1\columnwidth]{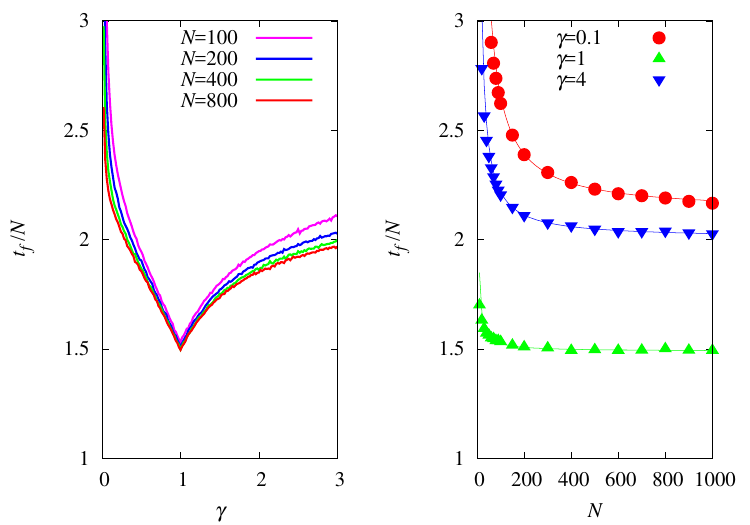}  
\caption{Scaled unconditional  mean fixation time $t_f/N$  as a function of $\gamma$ for  $N=100$, $200$, $400$ and $800$ (left panel).  The right panel shows $t_f/N$ as a function of $N$ for $\gamma=0.1$, $1$ and $4$.  The curves are the fit function $t_f/N =a + b/N$, where $a$ and $b$ are fit parameters.The initial conditions are $x_0=y_0=0.5$.
 }  
\label{fig:8}  
\end{figure}

As discussed above, the effects of the initial conditions are negligible for very large $N$, so we will only consider symmetric initial conditions $x_0=y_0=0.5$. The results are shown in Fig. \ref{fig:8}.   As expected, fixation is faster for $\gamma=1$ because this is the only case where the more successful models are copied with certainty, i.e. the probabilities (\ref{probs}) and  (\ref{probr}) are both equal to 1. This amounts to a greedy update strategy that leads to quick fixation.   More importantly, the right panel of the figure shows that $t_f$ increases linearly with large  increasing $N$, so the scaled fixation time $t_f/N$  tends to a non-zero finite value in the limit $N\to \infty$. This linear scaling of the mean fixation time with  $N$  contrasts remarkably with the half-lives of coexistence fixed points, which  grow exponentially with $N$ away from the threshold region  and as a typically small power of $N$ at the threshold \cite{Antal_2006,Fontanari_2024a}.

\section{Birth-death process}\label{sec:bd}

Here we use results from the standard birth-death process \cite{Karlin_1975} to derive equations for the fixation probability and conditional mean fixation times.   The Monte Carlo  study of these quantities  was presented in the previous section. 

Assume that at time $t=0$ the population consists of $X$ liars and $Y$ believers. We want to calculate   the probability $p_{lb}(X,Y)$ that the liars fixate in the sender population and the believers fixate in the receiver population.
 The key quantities needed for this calculation are the probabilities $T_X^+(X,Y)$ that $X$ increases by one and $Y$ is unchanged in time step $\delta t$, $T_X^-(X, Y)$ that $X$ decreases by one and $Y$ is unchanged in time step $\delta t$, $T_Y^+(X,Y)$ that $Y$ increases by one and $X$ is unchanged in time step $\delta t$, and $T_Y^-(X,Y)$ that $Y$ decreases by one and $X$ is unchanged in time step $\delta t$. We recall that $\delta t = 1/N$. It is clear that it will never happen that both $X$ and $Y$ will change in the same time step $\delta t$: in the game between the focal sender and the focal receiver, one of them will necessarily win, and the winner will not copy the model playing her role. According to  the imitation dynamics described in Section \ref{sec:id} we have
 \begin{equation}\label{Txp}
 T_X^+(X,Y) = \left ( \frac{ N-X}{N} \frac{Y}{N} \right )  \left (  \frac{ X}{N-1} \frac{Y-1}{N-1}   \right )  \left ( \frac{b_s+c_s}{\Delta f_{max}} \right ) ,
 \end{equation}
 where $\Delta f_{max} $ is given in eq.  (\ref{Dmax}).
Here, the term in the first  brackets gives the probability of selecting a truth teller from the $N$ senders and a believer from the $N$ receivers as focal players.  Note that the only way to increase the number of liars by one is to select a truth teller as the focal sender player. The payoff of the focal  truth teller is $-c_s$. The term in 
the second brackets gives the probability of selecting a liar  from the $N-1$ senders  and a different  believer from the $N-1$ receivers as model  players. The payoff of the model  liar is $b_s$. The term in  the third brackets is the probability that the focal sender will copy the model sender.   The other transition probabilities are
 \begin{eqnarray}
 T_X^-(X,Y)   & = &  \left ( \frac{X}{N} \frac{N-Y}{N} \right )  \left (  \frac{ N-X}{N-1} \frac{N-Y-1}{N-1}   \right )  \left ( \frac{b_s+c_s}{\Delta f_{max}} \right ) ,  \label{Txm} \\
 T_Y^+(X,Y)   & = &  \left (  \frac{N-Y}{N} \frac{N-X}{N} \right )  \left (  \frac{Y}{N-1} \frac{N-X-1}{N-1}   \right )  \left ( \frac{b_r+c_r}{\Delta f_{max}} \right )  , \label{Typ} \\
 T_Y^-(X,Y)   & = &  \left (  \frac{Y}{N} \frac{X}{N} \right )  \left (  \frac{N-Y}{N-1} \frac{X-1}{N-1}   \right )  \left ( \frac{b_r+c_r}{\Delta f_{max}} \right )  ,  \label{Tym}
\end{eqnarray}
which can be easily deduced from the explanation of the terms that are part of eq.  (\ref{Txp}). For ease of notation, it is convenient to define the probability that a change will occur at time step $\delta t$,
\begin{equation}\label{Ttot}
T (X,Y) = T^+_X (X,Y) + T^-_X (X,Y) + T^+_Y (X,Y)  +  T^-_Y (X,Y) .
\end{equation}
Given the transition probabilities, we can immediately write down the set of equations that determine $p_{lb}(X,Y)$ \cite{Karlin_1975},
 \begin{eqnarray}\label{elb0}
 p_{lb} (X,Y) & = & T^+_X (X,Y)  p_{lb}(X+1,Y) +  T^-_X (X,Y)  p_{lb} (X-1,Y) \nonumber \\
 &  & + T^+_Y (X,Y)  p_{lb} (X,Y+1)+  T^-_Y (X,Y)  p_{lb}(X,Y-1) \nonumber \\
 & & + \left [  1 -T(X,Y) \right ]   p_{lb} (X,Y)  
 \end{eqnarray}
with $X, Y=1,\ldots,N-1$. The boundary conditions are $p_{lb} (0,Y) = p_{lb} (X,0) =   0$  for all $X$ and $Y$,  $p_{lb} (X,N) =1$  for $X\neq 0$, $p_{lb} (N,Y) =0$ for $Y\neq N$.   This equation can be rewritten as 
 \begin{eqnarray}\label{elb1}
 p_{lb} (X,Y) & = & \frac{1}{T(X,Y)}  \left [ T^+_X (X,Y)  p_{lb}(X+1,Y) +  T^-_X (X,Y)  p_{lb} (X-1,Y) \right. \nonumber \\
 &  &  \left. + T^+_Y (X,Y)  p_{lb} (X,Y+1)+  T^-_Y (X,Y)  p_{lb}(X,Y-1)  \right ]
 \end{eqnarray}
 which can easily be solved self-consistently by setting the initial guess as   $p_{lb} (X,Y)  \break \sim \mbox{Uniform} (0,1)$ for  $X, Y=1,\ldots,N-1$ and iterating until convergence.  When applied to eq. (\ref{elb0}), however,  the iteration  quickly leads to divergence. The stability of the iteration procedure using eq. (\ref{elb1}) is due to the fact that the ratio on the right side of this equation is always less than 1.
 
The probabilities of the dynamics freezing in the other absorbing states, i.e. $p_{ld}(X,Y)$, $p_{tb}(X,Y)$ and $p_{td}(X,Y)$, given $X$ liars and $Y$ believers at time $t=0$ are also given by eqs. (\ref{elb0}) and (\ref{elb1}) with the appropriate change of subscripts, but the boundary conditions are different, as described next.
 \begin{enumerate}[label=(\alph*)]
 
 \item In the case  of fixation of liars and disbelivers, we have $p_{ld} (0,Y) = p_{ld} (X,N) \break =   0$  for all $X$ and $Y$,  $p_{ld} (N,Y) =1$  for $Y\neq N$, and $p_{ld} (X,0) =0$ for $ X\neq N$.
 
 \item In the case  of fixation of truth tellers  and believers, we have $p_{tb} (N,Y) = p_{tb} (X,0) =   0$  for all $X$ and $Y$,  $p_{tb} (0,Y) =1$  for $Y\neq 0$, and $p_{tb} (X,N) =0$ for $ X\neq 0$.
 
 \item  In the case  of fixation of truth tellers  and disbelivers, we have $p_{td} (N,Y) = p_{td} (X,N) =   0$  for all $X$ and $Y$,  $p_{td} (0,Y) =0$ for $ Y \neq 0$, and $p_{td} (X,0) =1$  for $X\neq N$.
 
 \end{enumerate}

We now turn to the calculation of the conditional mean fixation times, e.g.,   the mean time for the dynamics to freeze in the absorbing state $X=N$ and $Y=N$, considering only orbits that fall into this state, which we denote by $t_{lb}$.  Let $P_{lb} (t|X,Y)$ be the probability that the fixation of liars and believers occurs at time $t$ if there were $X$ liars and $Y$ believers at time $t=0$. We recall that the  time step $\delta t=1/N$ is nonzero for finite $N$, so  time is a discrete variable that takes on the values $t= 0, 1/N, 2/N, \ldots$.  We have
 \begin{eqnarray}\label{Plbt}
 P_{lb} (t|X,Y) & = &  \left [ T^+_X (X,Y)  P_{lb}(t' |X+1,Y) +  T^-_X (X,Y)  P_{lb} (t' |X-1,Y) \right. \nonumber \\
 &  &   \left. + T^+_Y (X,Y)  P_{lb} (t' |X,Y+1)+  T^-_Y (X,Y)  P_{lb}(t'|X,Y-1)  \right ] \nonumber \\
 & &  + \left [  1 - T(X,Y) \right ]   P_{lb} (t'|X,Y) 
 \end{eqnarray}
 where $t'= t-\delta t  = t- 1/N$.   
Summing over all possible fixation times yields the ultimate fixation probability
\begin{equation}
p_{lb}(X,Y) = \sum_{t=0}^\infty P_{lb} (t|X,Y) .
\end{equation}
The conditional mean time to fixation of liars and believers is defined as
\begin{equation}
t_{lb}(X,Y) = \frac{1}{p_{lb}(X,Y)}\sum_{t=0}^\infty t P_{lb} (t|X,Y).
\end{equation}
A  trite calculation (see, e.g., \cite{Antal_2006}) leads to the set of equations
 \begin{eqnarray}\label{tlb1}
 \tau_{lb} (X,Y) & = & \frac{1}{T(X.Y)}  \left [ T^+_X (X,Y)   \tau_{lb}(X+1,Y) +  T^-_X (X,Y)   \tau_{lb} (X-1,Y) \right. \nonumber \\
 &  & + T^+_Y (X,Y)  \tau_{lb} (X,Y+1)+  T^-_Y (X,Y)   \tau_{lb}(X,Y-1)     \nonumber \\
 & & \left. + \frac{1}{N} p_{lb} (X,Y)  \right ]
 \end{eqnarray}
 for $X=1,\ldots, N-1$ and $Y=1,\ldots, N$. Here $\tau_{lb} (X,Y)  =  p_{lb} (X,Y)  t_{lb}(X,Y) $ obeys the boundary conditions $\tau_{lb} (0,Y) = \tau_{lb} (X,0) =   0$  for all $X$ and $Y$,   and  $\tau_{lb} (N,Y) =0$ for all $Y$. In this last condition we have  $\tau_{lb} (N,Y) =0$ for $Y \neq N$ because  $p_{lb} (N,Y) =0$, but  $\tau_{lb} (N,N) =0$ because $t_{lb} (N,N) =0$. Note that $\tau_{lb} (X,N)$  with  $X=1,\ldots,N-1$ is not given by the boundary conditions and so must obtained from eq. (\ref{tlb1}).  In addition, since $T^+_Y (X,N) = 0$,  it is not necessary to define $\tau_{lb} (X,M+1)$ in  eq. (\ref{tlb1}).  Starting from random $\tau_{lb} (X,Y)$ for $0<X<N$ and  
$ 0<Y\leq N$, this equation can easily be solved self-consistently once $ p_{lb} (X,Y)$ is known. 

\begin{figure}[th] 
\centering
 \includegraphics[width=1\columnwidth]{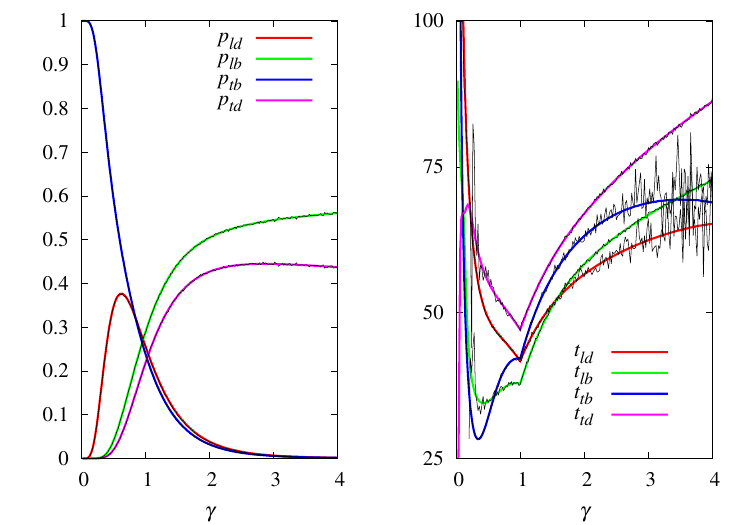}  
\caption{Probability of fixation (left panel) and mean conditional fixation time (right panel) of the sender and receiver  strategies as a function of $\gamma$ for  $x_0=0.5$ and  $y_0=0.1$.  The number of players in each role is   $N=50$. The colored thick curves  are the solutions of the birth-death equations and the black thin curves are the results of the Monte Carlo simulations. 
 }  
\label{fig:9}  
\end{figure}
 
We  note that $\tau_{ld}(X,Y)$, $\tau_{tb}(X,Y)$, and $\tau_{td}(X,Y)$ are given by eq. (\ref{tlb1})  by setting the subscripts properly for each case.  What is different are the ranges of $X$ and $Y$ to which the counterparts of eq. (\ref{tlb1}) are applied and the boundary conditions. Next, we consider each case separately.

 \begin{enumerate}[label=(\alph*)]
 
 \item In the case  of fixation of liars and disbelivers, we have $\tau_{ld} (0,Y) = \tau_{ld} (X,N)  \break =  \tau_{ld} (X,0)=  0$  for all $X$ and $Y$.  The counterpart to eq.  (\ref{tlb1}) is applied for   $X=1,\ldots, N$ and $Y=1,\ldots, N-1$.
 
 \item In the case  of fixation of truth tellers  and believers, we have $\tau_{tb} (N,Y) = \tau_{tb} (X,0) =  \tau_{tb} (X,N) =0$  for all $X$ and $Y$. The counterpart to eq.  (\ref{tlb1}) is applied for   $X=0,\ldots, N-1$ and $Y=1,\ldots, N-1$.
 
 \item  In the case  of fixation of truth tellers  and disbelivers, we have $\tau_{td} (N,Y) = \tau_{td} (X,N) =  \tau_{td} (0,Y)= 0$  for all $X$ and $Y$. The counterpart to eq.  (\ref{tlb1}) is applied for   $X=1,\ldots, N-1$ and $Y=0,\ldots, N-1$.
 
 \end{enumerate}

 In Fig. \ref{fig:9} we show the solutions of the self-consistent method to solve the birth-death equations together with the results of the Monte Carlo simulations. The fixation probabilities are indistinguishable between the two approaches, as are the mean conditional fixation times when the number of fixations in the absorbing state is sufficiently large. For example, the large fluctuations in $t_{tb}$ for $\gamma >2$ are a consequence of the very small number of fixations of truth tellers and believers in this region.
 
We emphasize, however, that the birth-death approach is not a substitute for Monte Carlo simulations.  This is because the iteration of the self-consistent method to solve eqs. (\ref{elb1}), (\ref{tlb1}) and their counterparts for the other absorbing states requires high numerical precision, and the accumulation of numerical errors becomes unmanageable as $N$ increases. This also happens in the cases where it is possible to derive explicit expressions for the fixation probability and the mean conditional fixation times in terms of the transition probabilities \cite{Karlin_1975,Antal_2006}: these expressions involve the sums of products, which quickly become unmanageable to evaluate numerically as $N$ increases. Nevertheless, the birth-death approach is the only way to validate the Monte Carlo results for not too large $N$, and the excellent agreement between the two approaches shown in Fig. \ref{fig:9} supports the analysis presented in Section \ref{sec:mc}.

\section{Discussion}\label{sec:disc}

A significant advantage of both the deterministic and stochastic versions of our model is the robustness of our results to the specific values of the payoff parameters $b_s$, $c_s$, $b_r$ and $c_r$. Instead, the dynamics are governed solely by the ratio $\gamma$, defined in eq. (\ref{gamma}). While directly mapping these simplified model parameters to precise real-world scenarios presents challenges, we can offer a clear and insightful interpretation for the ratio $\gamma$. Specifically, $\gamma > 1$ signifies that receivers exhibit a higher evolutionary rate and greater stakes than senders, whereas $\gamma < 1$ indicates the converse.

The finite population analysis of the sender-receiver game yields markedly different results compared to the replicator equation framework  \cite{Gaunersdorfer_1991,Hofbauer_1998}.  The replicator equation, which applies in the deterministic limit ($N \to \infty$), predicts oscillatory behavior. This oscillation arises from the cyclical dominance of strategies in each role: for example, a temporary increase in liars among senders promotes the growth of disbelievers among receivers (see Fig. \ref{fig:1}).  However, our finite-size scaling analysis reveals a sharp threshold in the same limit ($N \to \infty$):  receivers are guaranteed to win if senders evolve more rapidly ($\gamma < 1$), and senders are guaranteed to win if receivers evolve more rapidly ($\gamma > 1$).  At the critical threshold $\gamma_c=1$ neither role has an advantage.  This leads to the surprising conclusion that the slower evolving role ultimately prevails, because it can exploit the fixed strategy of the faster-evolving role.   The evolutionary speeds of the sender and receiver roles are given by  the constants $(b_s+c_s)/\Delta f_{max}$ and  $(b_r+c_r)/\Delta f_{max}$, respectively, which appear in the replicator equations (\ref{x_rep}) and (\ref{y_rep}).

The stark difference in long-term evolutionary outcomes stems from the order in which the limits $t \to \infty$ and $N \to \infty$ are evaluated. 
The replicator equation's predictions are recovered from the stochastic imitation dynamics by first fixing time $t$ and then taking the population size to infinity ($N \to \infty$), as illustrated in Fig. \ref{fig:3}. Conversely, our analysis of fixation probabilities and mean fixation times involves fixing the population size $N$, allowing the dynamics to run to an absorbing state ($t \to \infty$), and then repeating this process for increasingly large $N$ to extrapolate to the $N \to \infty$ limit (Figs. \ref{fig:6} and \ref{fig:7}). Given that real-world populations are finite and long-term predictions are the primary concern, the latter approach is more relevant.  This finite population study employed an analytical birth-death process approach, which becomes computationally challenging for large $N$, and extensive Monte Carlo simulations combined with finite-size scaling, which  proved to be a powerful method for analyzing the stochastic dynamics of large populations.

We find not only that the probability of the slowest evolving role winning increases with population size and tends to unity as $N \to \infty$, but also a more surprising result: in finite populations, the initially less frequent strategy within the slower role is more likely to become fixed.  Consequently, the slowest and rarest strategies are favored.  However, as $N$ increases, the impact of initial frequencies lessens, because the stochastic dynamics explore a broader range of frequencies before settling into an absorbing state.  This contrasts sharply with the replicator equation framework, where initial frequencies dictate the closed trajectory in the phase plane.

A striking result of our study is the linear scaling of mean fixation time with population size $N$.  This contrasts markedly with the exponential scaling of half-life seen in metastable coexistence fixed points of replicator equations \cite{Antal_2006,Fontanari_2024a}.  Although some conservative systems, such as the Lotka-Volterra model, are robust to demographic noise and can maintain (non-periodic) oscillations through coherence resonance even in very large populations \cite{McKane_2005}, the sender-receiver game's rapid escape from deterministic orbits (Fig. \ref{fig:4}) prevents these orbits from being considered metastable states of the stochastic dynamics.

While our analysis provides insights into the evolutionary dynamics of deception in a well-mixed population, real-world interactions are often structured by social networks \cite{Granovetter_1973}. Investigating the impact of network topology on our sender-receiver game represents a natural and important extension. For instance, in scale-free networks, highly connected individuals could play a disproportionate role in the spread of deceptive strategies, whereas in small-world networks, local clustering could lead to the formation of strategy clusters, influencing the overall evolutionary outcome \cite{Newman_2018}. 
Future research could explore these effects by incorporating network structure into our model. This, however, poses a particularly complex challenge, as it necessitates the introduction of three, in principle, distinct networks forming a multilayer network \cite{Domenico_2013}. The first defines the undirected links between senders, dictating the imitation dynamics among them. The second similarly defines the undirected links between receivers, governing their imitation dynamics. Finally, the third is a directed network connecting senders to receivers, representing the flow of information.

Our study of deception in a sender-receiver game intersects with broader research on information diffusion and opinion dynamics. Indeed, evolutionary game theory has been fruitfully applied to these areas \cite{Jiang_2014,Li_2022,Soares_2024}.  In particular, studies of information diffusion in structured populations highlight how factors such as public knowledge, interaction feedback, and network structure influence the spread of competing opinions or information. However, our approach differs significantly in  at least two key aspects. First, agents in our model have distinct and fixed roles (senders and receivers), meaning that there is no information flow between agents with the same role, as seen in typical information diffusion models. In our model, information flows only from sender to receiver.  Second, the unique asymmetric payoff structure of our model leads to cyclical behavior in the infinite population limit, in contrast to the equilibrium scenario typically observed in information diffusion models.

In conclusion, our finite-population analysis of this sender-receiver game, which models the strategic interplay between deceptive senders and discerning receivers, reveals a counterintuitive outcome: the role with the larger difference in payoffs between successful and unsuccessful interactions (i.e., $b_s+c_s$ for senders and $b_r+c_r$ for receivers) is ultimately more susceptible to defeat. This suggests that over-reliance on high-stakes deception can be counterproductive in the long run, favoring instead subtle, low-risk manipulation strategies for sustained influence. Indeed, numerous manipulative social media tactics exemplify these principles. For example, content designed for algorithmic amplification, such as highly emotional or divisive material, minimizes risk by leveraging algorithms for dissemination. Similarly, the creation of fictitious online accounts for spreading disinformation offers reduced visibility and diminishes the risk of direct exposure. Moreover, the gradual erosion of democratic norms and the normalization of extremist rhetoric are facilitated by incrementally introducing extreme or controversial concepts, beginning with less radical iterations, thereby ensuring that each step elicits a less pronounced reaction.

\section*{Acknowledgments}
JFF is partially supported by  Conselho Nacional de Desenvolvimento Cient\'{\i}fico e Tecnol\'ogico  grant number 305620/2021-5. EVMV is supported by  Conselho Nacional de Desenvolvimento Cient\'{\i}fico e Tecnol\'ogico  grant number 131817/2023-0.

\appendix

\renewcommand{\theequation}{A.\arabic{equation}}
\setcounter{equation}{0}
\setcounter{figure}{0}

\section{}\label{ref:A}

Here  we show how the period for  small oscillations given in eq. (\ref{Ts2}) can be derived from the general period equation (\ref{Tg1}) by taking the  limits  $x_0 \to 1/2$ and $y_0 \to 1/2$ or, equivalently, the limit $C \to  1/4^{1+\gamma}$. Introducing the new variable
\begin{equation}
\epsilon = 1/4^{1+\gamma} - C \ll 1
\end{equation}
allows us to write the integration limits in eq. (\ref{Tg1}) as $x_{min} \approx  1/2 - 4^{\gamma/2} (\epsilon/\gamma)^{1/2}$ and  $x_{max} \approx 1/2 + 4^{\gamma/2} (\epsilon/\gamma)^{1/2}$. It is now clear that  the integration variable $x$  in eq. (\ref{Tg1}) varies only in the close vicinity of $1/2$, so we can write this equation as 
\begin{equation}
\tau \approx  4^{3/2-\gamma/2}  \int_{1/2 - 4^{\gamma/2} (\epsilon/\gamma)^{1/2}}^{1/2 + 4^{\gamma/2} (\epsilon/\gamma)^{1/2}} \frac{dx}{\sqrt{ [x(1-x)]^\gamma - 1/4^{\gamma} +  4 \epsilon}} .
\end{equation}
Changing the integration variable $z = 1/2 -x$ so that $x(1-x) = 1/4-z^2$ yields
\begin{equation}
\tau \approx  4^{2-\gamma/2}  \int_{0}^{4^{\gamma/2} (\epsilon/\gamma)^{1/2}} \frac{dz}{\sqrt{ (1/4-z^2)^\gamma - 1/4^{\gamma} +  4 \epsilon}} .
\end{equation}
Since $z < 4^{\gamma/2} (\epsilon/\gamma)^{1/2} \ll 1$ we can write $(1/4 -z^2)^\gamma \approx 1/4^{\gamma}  - 4^{1-\gamma} \gamma z^2$, so that
\begin{eqnarray}
\tau  & \approx  & 4^{2-\gamma/2}  \int_{0}^{4^{\gamma/2} (\epsilon/\gamma)^{1/2}} \frac{dz}{ \sqrt{4 \epsilon - 4^{1-\gamma} \gamma z^2}}  \\
& \approx  & 4^{3/2} \gamma^{-1/2}  \int_{0}^{4^{\gamma/2} (\epsilon/\gamma)^{1/2}} \frac{dz}{ \sqrt{  4^\gamma \epsilon/\gamma -z^2}}  \label{int1}\\
& \approx  & 4^{3/2} \gamma^{-1/2} \int_0^{\pi/2} d\theta = 4 \pi \gamma^{-1/2},    \label{int2}
\end{eqnarray}
where we have changed the variable $z = 4^{\gamma/2} (\epsilon/\gamma)^{1/2}  \cos \theta$ to go from  eq. (\ref{int1}) to eq. (\ref{int2}).

\section{}\label{ref:B}
\renewcommand{\theequation}{B.\arabic{equation}}
\setcounter{equation}{0}
\setcounter{figure}{0}

Here we show how to remove the divergences at $x_{min}$ and $x_{max}$, eqs. (\ref{xmin}) and   (\ref{xmax}), of the integrand in eq.  (\ref{Tg1}). We rewrite eq.  (\ref{Tg1}) as
\begin{equation}
\tau =  2(A+B),
\end{equation}
where
\begin{equation}\label{A}
A= \int_{x_{min}}^{1/2} \frac{dx}{[x(1-x)]^{1-\gamma/2} \sqrt{ [x(1-x)]^\gamma -4 C}}
\end{equation}
and
\begin{equation}\label{B}
B= \int_{1/2}^{x_{max}} \frac{dx}{[x(1-x)]^{1-\gamma/2} \sqrt{ [x(1-x)]^\gamma -4 C}}, 
\end{equation}
since $x_{min} \leq 1/2 \leq x_{max}$.  The symmetry of the trajectories shown in Fig. \ref{fig:1} indicates that $A=B$, but we can easily prove this equality by  changing  the integration variable to $ z =x(1-x)$ and  noting that $dz = (1-2x)dx = \sqrt{1-4z} dx$ in eq. (\ref{A}) because $x \leq 1/2$, and $dz = (1-2x)dx = - \sqrt{1-4z} dx$ in eq. (\ref{B}) because $x \geq 1/2$.  The result is
\begin{equation}\label{AA}
A =  B= \int_{(4C)^{1/\gamma}}^{1/4} \frac{dz}{z^{1-\gamma/2} \sqrt{ (1-4z)(z^\gamma -4 C)}}
\end{equation}
since $x_{min}(1-x_{min}) = x_{max}(1-x_{max}) = (4C)^{1/\gamma}$. 
Unfortunately, the integrand still diverges at both integration extremes, forcing us to split the integral again, $A = A_1+A_2$, where
\begin{equation}\label{A1}
A_1= \int_{(4C)^{1/\gamma}}^{a} \frac{dz}{z^{1-\gamma/2} \sqrt{ (1-4z)(z^\gamma -4 C)}}
\end{equation}
and
\begin{equation}\label{A2}
A_2= \int_{a}^{1/4} \frac{dz}{z^{1-\gamma/2} \sqrt{ (1-4z)(z^\gamma -4 C)}}.
\end{equation}
Here $a = [(4C)^{1/\gamma}+1/4]/2$ is the midpoint of the integration interval in eq. (\ref{AA}).
Changing the integration variable in eq.  (\ref{A1}) to $u = \sqrt{z^\gamma -4C}$  yields the  proper integral
\begin{equation}\label{A11}
A_1 =  \frac{2}{\gamma} \int_{0}^{\sqrt{a^\gamma - 4C}} \frac{du}{\sqrt{ (u^2 + 4C)[1-4(u^2 + 4C)^{1/\gamma}]}}. 
\end{equation}
Similarly, changing the integration variable in eq.  (\ref{A2}) to $u = \sqrt{1-4z}$  yields
\begin{equation}\label{A21}
A_2 =2 \int_{0}^{\sqrt{1-4a}} \frac{du}{(1-u^2)^{1-\gamma/2} \sqrt{ (1-u^2)^\gamma -4^{1+\gamma} C}}, 
\end{equation}
which is also a proper integral. Thus, we have removed the integrand divergences from eqs. (\ref{A}) and  (\ref{B}). The integrals (\ref{A11}) and (\ref{A21}) can easily be  evaluated numerically using standard integration routines \cite{Press_1992}. The period for oscillations of  general amplitude  is then
\begin{equation}
\tau =  4(A_1+A_2) ,
\end{equation}
which is shown in Fig. \ref{fig:2} for a representative selection of  model parameters.



\begin{thebibliography}{999}



\bibitem{Plato_2007}
Plato.:
The Republic.
Penguin Books, 2nd ed., New York (2007)

 \bibitem{Kant_2012}
Kant, I.:
Groundwork of the Metaphysics of Morals.
 Cambridge University Press, Cambridge, UK (2012)
 
\bibitem{Sober_1994}
 Sober, E.:
 The primacy of truth-telling and the evolution of lying. 
 In: Sober, E. (ed.)  From a Biological Point of View: Essays in Evolutionary Philosophy,
  pp. 71--92.  Cambridge University Press, Cambridge, UK (1994)
  
\bibitem{Fontanari_2023}
 Fontanari, J.F.:
Kant's Modal Asymmetry between Truth-Telling and Lying Revisited.
Symmetry {\bf 15}  555  (2023) https://doi.org/10.3390/sym15020555

\bibitem{Maynard_2003}
 Maynard Smith, J., Harper, D.:
 Animal Signals.
 Oxford University Press,  Oxford, UK (2003)
 
 \bibitem{Zahavi_1975}
 Zahavi, A.:
 Mate selection-A selection for a handicap. 
   J. Theor. Biol. {\bf 53} 205--214 (1975) 
   https://doi.org/10.1016/0022-5193(75)90111-3

\bibitem{Fallis_2015}
Fallis, D.:
 What Is Disinformation? 
 Libr. Trends  {\bf 63} 401--426 (2015)
 https://doi.org/10.1353/lib.2015.0014

\bibitem{Seger_2020}
Seger, E., Avin, S.,  Pearson, G., Briers, M.,  Heigeartaigh, S.O.,  Bacon, H.:
 Tackling Threats to Informed Decision-Making in Democratic Societies.
 The Alan Turing Institute: London, UK  (2020) 
 https://www.turing.ac.uk/news/publications/tackling-threats-informed-decision-making-democratic-societies
 
  \bibitem{Maynard_1982}
 Maynard Smith, J.:
 Evolution and the Theory of Games.
 Cambridge University Press, Cambridge, UK (1982)
 
\bibitem{Kreps_1994}
Kreps D., Sobel J.:
Signalling.
In: Aumann R., Hart S. (eds.), The Handbook of Game Theory, Volume II,  pp. 849--867.
North-Holland, Amsterdam (1994) 
 
 \bibitem{Erat_2012}
  Erat, S.,  Gneezy, U.:
  White lies. 
  Manag. Sci. {\bf 58} 723--733 (2012)
  https://doi.org/10.1287/mnsc.1110.1449
  
 \bibitem{Capraro_2019}
  Capraro, V., Perc, M., Vilone, D.: 
  The evolution of lying in well-mixed populations. 
  J. R. Soc. Interface {\bf 16} 20190211 (2019)
  https://doi.org/10.1098/rsif.2019.0211
  
 \bibitem{Capraro_2020} 
Capraro, V., Perc, M., Vilone, D.:
 Lying on networks: The role of structure and topology in promoting honesty. 
 Phys. Rev. E  {\bf 101} 032305 (2020)
 https://doi.org/10.1103/PhysRevE.101.032305
 
 \bibitem{Gaunersdorfer_1991}
 Gaunersdorfer, A., Hofbauer,  J.,   Sigmund, K.: 
 On the dynamics of asymmetric games.
 Theor. Popul. Biol. {\bf 39} 345--357 (1991)
https://doi.org/10.1016/0040-5809(91)90028-E
 
\bibitem{Hofbauer_1998}
Hofbauer, J., Sigmund, K.:
 Evolutionary Games and Population Dynamics.
 Cambridge University Press, Cambridge, UK (1998)
 
 \bibitem{Traulsen_2005}
Traulsen, A.,  Claussen, J.C.,  Hauert, C.: 
Coevolutionary dynamics: From finite to infinite populations.
 Phys. Rev. Lett. {\bf 95}  238701 (2005)
  https://doi.org/10.1103/PhysRevLett.95.238701
  
 \bibitem{Ohtsuki_2006}
 Ohtsuki, H.,   Nowak,  M. A.:
 The Replicator Equation on Graphs.
 J. Theor. Biol. {\bf 243} 86--97 (2006)
https://doi.org /10.1016/j.jtbi.2006.06.004
  
  \bibitem{Fontanari_2024b}
Fontanari, J.F.:
 Imitation dynamics and the replicator equation.
Europhys. Lett. {\bf 146} 47001  (2024)
https://doi.org/10.1209/0295-5075/ad473e

 \bibitem{Newman_1999}
  Newman, M.E., Barkema, G.T.:
  Monte Carlo methods in statistical physics.
   Oxford University Press, Oxford, UK (1999)
 
 \bibitem{Landau_2000}
 Landau, D.P., Binder, K.: 
  A Guide to Monte Carlo Simulations in Statistical Physics.
  Cambridge University Press, Cambridge, UK (2000)
  
\bibitem{Privman_1990}
 Privman, V.:
  Finite-Size Scaling and Numerical Simulations of Statistical Systems.
  World Scientific, Singapore (1990)
  
 
  
  \bibitem{Karlin_1975}
Karlin, S.,  Taylor, H.M.:
A first course in stochastic processes.
Academic Press, New York (1975)

\bibitem{Antal_2006}
Antal, T.,   Scheuring, I.:
Fixation of Strategies for an Evolutionary Game in Finite Populations.
Bull. Math. Biol.  \textbf{68} 1923--1944  (2006)
https://doi.org/10.1007/s11538-006-9061-4

\bibitem{Fontanari_2024a}
Fontanari, J.F.:
Cooperation in the face of crisis: effect of demographic noise in collective-risk social dilemmas.
Math. Biosci. Eng.  {\bf 21} 7480--7500 (2024)  
https://doi.org/10.3934/mbe.2024329

\bibitem{Eigen_1971}
Eigen, M.:
Selforganization of matter and the evolution of biological macromolecules. 
Naturwissenschaften  {\bf 58}  465--526 (1971)
 https://doi.org/10.1007/BF00623322
 
 \bibitem{Mariano_2024}
 Mariano, M.S, Fontanari, J.F.:
 Evolutionary Game-Theoretic Approach to the Population Dynamics of Early Replicators.
 Life  {\bf 14} 1064 (2024)
 https://doi.org/10.3390/life14091064
 
 \bibitem{Silva_2009}
De Silva, H., Sigmund, K. :
 Public Good Games with Incentives: The Role of Reputation. 
 In: Levin, S.A. (ed). Games, Groups, and the Global Good, pp. 85--103. 
 Springer, New York (2009) 
 
  \bibitem{Vieira_2024}   
  Vieira, E.V.M., Fontanari, J.F.: 
  A Soluble Model for the Conflict between Lying and Truth-Telling.
  Mathematics  {\bf 12} 414 (2024)
 https://doi.org/10.3390/math12030414
 
 \bibitem{Britton_2003}
 Britton, N.F.:
Essential Mathematical Biology.
Springer,  London (2003)

\bibitem{Murray_2007}
Murray, J.D.:
Mathematical Biology: I. An Introduction.
Springer, New York (2007)

  \bibitem{Kirkpatrick_1994}
 Kirkpatrick, S.,  Selman, B.:
  Critical Behavior in the Satisfiability of Random Boolean Expressions.
Science {\bf 264}, 1297--1301 (1994)
   https://doi.org/10.1126/science.264.5163.1297

  \bibitem{Campos_1999}
 Campos, P.R.A.,  Fontanari, J.F.: 
 Finite-size scaling of the error threshold transition in finite populations.
  J. Phys. A Math. Gen. {\bf 32}  L1--L7 (1999)
   https://doi.org/10.1088/0305-4470/32/1/001
   

 
 \bibitem{McKane_2005}
 McKane,  A.J.,   Newman, T.J.:
 Predator-prey cycles from resonant amplification of demographic stochasticity.
Phys. Rev. Lett. {\bf 94}  218102 (2005) https://doi.org/10.1103/PhysRevLett.94.218102
 


\bibitem{Granovetter_1973}
Granovetter, M.: 
The strength of weak ties.
Am. J. Sociol. {\bf 78}, 1360--1380 (1973)
https://doi.org/10.1086/225469

\bibitem{Newman_2018}
Newman, M.E.:
Networks. 
 Oxford University Press, Oxford, UK (2018)

\bibitem{Domenico_2013}
De Domenico, M.,  Sol\'e-Ribalta, A.,  Cozzo, E.,  Kivel\"a, M., Moreno, Y.,  Porter, M.A., G\'omez, S.,  Arenas, A.:
Mathematical Formulation of Multilayer Networks.
Phys. Rev. X {\bf 3}, 041022 (2013)
https://doi.org/10.1103/PhysRevX.3.041022



\bibitem{Jiang_2014}
Jiang, C., Chen, Y., Liu, K.J.R.:
Evolutionary Dynamics of Information Diffusion Over Social Networks.
 IEEE Trans. Signal Process {\bf 62}, 4573--4586 (2014)
https://doi.org/10.1109/TSP.2014.2339799

\bibitem{Li_2022}
 Li, Z., Chen, X.,  Yang, H.-X., Szolnoki, A.:
Game-theoretical approach for opinion dynamics on social networks.
Chaos {\bf 32}, 073117 (2022) 
https://doi.org/10.1063/5.0084178

\bibitem{Soares_2024}
Soares, J.P.M., Fontanari, J.F.:
N-player game formulation of the majority-vote model of opinion dynamics.
Physica A  {\bf 643}, 129829 (2024)
https://doi.org/10.1016/j.physa.2024.129829
 
 \bibitem{Press_1992}
 Press, W.H., Teukolsky, S.A., Vetterling, W.T., Flannery, B.P.:
  Numerical Recipes in Fortran: The Art of Scientific Computing.
   Cambridge University Press, Cambridge, UK (1992)


\end{thebibliography}
\end{document}